%% file: main.tex
\documentclass[aps,prb,10pt,twocolumn, superscriptaddress, longbibliography, nobalancelastpage]{revtex4-2}

\input{preamble}

\begin{document}

\title{Spinon confinement and deconfinement in spin-1 chains}

\author{Laurens Vanderstraeten}
\affiliation{Department of Physics and Astronomy, University of Ghent, Krijgslaan 281, 9000 Gent, Belgium}

\author{Elisabeth Wybo}
\affiliation{Department of Physics and Astronomy, University of Ghent, Krijgslaan 281, 9000 Gent, Belgium}
\affiliation{Department of Physics and Institute for Advanced Study, Technical University of Munich, 85748 Garching, Germany}

\author{Natalia Chepiga}
\affiliation{Institute for Theoretical Physics, University of Amsterdam, Science Park 904, 1098 XH Amsterdam, The Netherlands}

\author{Frank Verstraete}
\affiliation{Department of Physics and Astronomy, University of Ghent, Krijgslaan 281, 9000 Gent, Belgium}

\author{Fr\'ed\'eric Mila}
\affiliation{Institute of Physics, Ecole Polytechnique F\'ed\'erale de Lausanne (EPFL), CH-1015, Lausanne, Switzerland}

\begin{abstract}
Motivated by the rich phase diagrams recently unveiled in frustrated spin-1 chains with competing next-nearest neighbour and four-spin interactions, we investigate the nature of the elementary excitations of spin-1 chains in the vicinity of phase transitions using the matrix-product state (MPS) representation of elementary excitations. First, we show that both spinons and magnons naturally arise in $\mathrm{SU}(2)$ invariant spin chains when describing ground states and elementary excitations using MPS. Then we investigate the nature of the elementary excitations across the first-oder transition between the Haldane phase and the topologically trivial next-nearest neighbor Haldane phase. We show explicitly that spinons deconfine at the transition, and we calculate the dispersion of these deconfined spinons with MPS. We also show that, immediately away from the transition line, spinons confine, forming dispersive spinon/anti-spinon bound states on both sides of the transition. Finally, we show that deconfined spinons also appear at the transition between the Haldane phase and the spontaneously dimerized phase when this transition is first order.
\end{abstract}

\maketitle

\section{Introduction}

In 1982, Faddeev and Takhtajan published a paper with the enigmatice title ``What is the spin of a spin wave?'' \cite{Faddeev1981, Faddeev1984}. In this paper, the authors showed that the elementary excitations in the spin-1/2 Heisenberg antiferromagnetic chain are spin-1/2 doublets, contrary to the common belief at that time that the spectrum should consists of triplet spin waves \cite{DesCloizeaux1962}. They consider these excitations to be particles---later, they were called spinons \cite{Haldane1988, Haldane1991}---by virtue of their localized nature and the fact that one can consider their scattering. All physical states, i.e. the states that can be created by acting with a local operator on the ground state, consist of an even number of these spinon states. Around the same time, Shastry and Sutherland proposed the spin-1/2 soliton to be the elementary excitation in dimerized spin-1/2 chains \cite{Shastry1981}, similar to the solitons in dimerized long-chain polymers \cite{Su1979}. Again, the solitons are considered to be localized spin-1/2 particles and only two-particle states connect to the ground state through a local operator.

\par Later on it was realized that adding extra terms in the spin-chain hamiltonian can lead to the confinement of these fractionalized particles. For example, adding an explicit dimerization to the dimerized spin chain leads to an effective linear potential between the solitons and, therefore, to their confinement \cite{Sorensen1998, Affleck1998, Augier1999}. With the spinons no longer existing as well-defined particles, the spectrum of such an extended model consists of a stack of spinon/anti-spinon bound states \cite{Shiba1980}. The phenomenon of spinon or soliton confinement has non-trivial effects in the real-time dynamics of spin chains as well \cite{Kormos2017, Liu2019}.

\par The situation is quite different for the Heisenberg spin-1 chain. As first shown by Haldane\cite{Haldane1983a, Haldane1983b} and further confirmed by the investigation of an extended model with biquadratic interactions at the special point known as the AKLT point \cite{Affleck1987, Affleck1988}, the spectrum of the spin-1 chain is gapped, and the low-lying excitations are gapped magnons. As for the spin-1/2 chain with explicit dimerization, these magnons can be interpreted as the result of confining spinons, a point of view emphasized by Greiter and collaborators in a series of papers dedicated to $\mathrm{SU}(N)$ chains\cite{Greiter2007,Greiter2009}. Spin-1/2 excitations appear at the edges of spin-1 chains of finite length, but they cannot appear as free excitations in the bulk of the chain.

\par In this paper, we show that deconfined spinons actually exist in $\mathrm{SU}(2)$ invariant spin-1 chains at the transition between the Haldane phase and another phase provided two conditions are met: the other phase does not have spin-1/2 edge excitations, and the transition is first order. The spinons are spin-1/2 excitations that appear at domain walls between the two types of phases, and as long as one sits on the first order transition line, they exist as gapped by deconfined excitations with a dispersion that we have calculated explicitly along two transitions for a model with next-nearest neighbor and three-site interactions\cite{Chepiga2016}.

\par This investigation is part of a more general program. Spinons in quantum spin chains were among the first instances of collective excitations with fractional quantum numbers, and are at the basis of many exciting developments in strongly-correlated quantum physics. In particular, it was gradually realized that fractionalized excitations are typically supported by a ground state that exhibits some form of topological order. This connection between topological order and fractionalized excitations can be naturally understood within the language of tensor networks. In this language, a quantum ground state is represented as the contraction of an extensive number of local tensors, where the topological properties of the global state are encoded as symmetries of the local tensors \cite{Schuch2010}. Fractionalized excitations are represented as either defects in the symmetry pattern of the ground-state tensors \cite{Vanderstraeten2016, ZaunerStauber2018b} or local perturbations with a non-trivial string of symmetry operations \cite{Schuch2010, Haegeman2015}. In both cases, they are naturally related to the topological properties of the ground state.

\par In this paper, we continue this program of understanding fractionalized quasiparticles within the language of tensor networks. In the first part of the paper [Sec.~\ref{sec:mps}] we explain how the formalism of uniform matrix product states (MPS) gives a natural description of both spinons and magnons in spin chains. Hereto we review the formalism \cite{McCulloch2002, Weichselbaum2012} for incorporating (non-abelian) symmetries in MPS. Although this description confirms the picture that spinons typically occur in half-integer spin chains, we show explicitly how they can emerge in a $\mathrm{SU}(2)$-invariant spin-1 chain at a first-order transition between a symmetry-protected topological (SPT) phase and a trivial phase. In the second part [Sec.~\ref{sec:spin1}], we provide numerical evidence for this scenario in a frustrated and dimerized spin-1 chain, and show how they are confined when tuning away from the transition.

\section{Spinons and matrix product states}
\label{sec:mps}

The formalism of translation-invariant matrix product states (MPS) in the thermodynamic limit---the so-called uniform MPS---has been developed for simulating static and dynamic properties of quantum spin chains \cite{Vanderstraeten2019}. In particular, it yields a natural description of elementary excitations as localized particles against a strongly-correlated background \cite{Vanderstraeten2015}. When implementing physical symmetries into the MPS parametrization, definite quantum numbers can be assigned to these particles \cite{ZaunerStauber2018b}. In this section, we explain how this formalism is applied to $\SU(2)$ spin chains and how particles with both integer and fractional quantum numbers naturally emerge from the MPS formalism.

\subsection{Ground states}

It is, by now, well-known that matrix product states (MPS) provide an efficient parametrization of ground states of (gapped) quantum spin chains \cite{Hastings2006, Verstraete2006}. Although most state-of-the-art MPS algorithms are formulated on finite chains \cite{Schollwoeck2011}, the MPS formalism is laid out most elegantly when working in the thermodynamic limit directly \cite{Vidal2007, Haegeman2011}. Indeed, a translation-invariant ground state can be represented as an MPS where we just repeat the same tensor $A$ on each site in the chain. This can be generalized to states with larger unit cells, where we repeat the same sequence of tensors $\{A_1, A_2, \dots\}$. The state is represented diagrammatically as
\begin{multline*}
\ket{\Psi(A_1,\dots,A_n)} \\= \dots \diagram{main}{1} \dots,
\end{multline*}
and is translation invariant over $n$ sites by construction. In recent years, it was shown that it is possible to variationally optimize over this set of states directly in the thermodynamic limit to find accurate ground-state approximations for a given hamiltonian \cite{McCulloch2008, ZaunerStauber2018a, Vanderstraeten2019}.
\par The real power of MPS is laid bare when imposing symmetry constraints on the tensors that reflect the physical symmetries in the system \cite{McCulloch2002, Weichselbaum2012}. Indeed, it has been realized that an MPS can only be invariant under certain global symmetry operations on the physical degrees of freedom if the virtual legs transform according to a representation of the same symmetry group \cite{PerezGarcia2006}. In the case of a one-site unit cell, if an MPS is invariant under the global symmetry operation $U(g)=\bigotimes_iu_i(g)$,
\begin{equation*}
U(g) \ket{\Psi(A)} = \ket{\Psi(A)}, \qquad \forall g,
\end{equation*}
it follows that the MPS tensor $A$ itself transforms as
\begin{equation*}
\diagram{main}{2} = \diagram{main}{3}, \qquad V_g\dag V_g = \one
\end{equation*}
Indeed, in the global MPS wavefunction the symmetry operations on the virtual legs, $V(g)$ and $V(g)\dag$, cancel such that the global state is invariant. In general, the representation $V_g$ can be decomposed in a direct sum of (projective) irreps of the physical symmetry group, so that the MPS tensor decomposes into a number of blocks that are labeled by the irreps on each leg. In order for the total MPS wavefunction to transform trivially under the global symmetry operation, it is required that the tensor itself only contains non-zero blocks for which the three irreps fuse to the trivial representation---i.e., the tensor itself globally transforms trivially. This property generalizes to larger unit cells, where the representations $V_g$ can be site-dependent within the unit cell.
\par In the case of a quantum spin chain with $\SU(2)$ invariance, where the physical degrees of freedom transform according to a specific spin-$s$ representation, the virtual degrees of freedom transform as a direct sum of representations of $\SU(2)$ labeled by $j_1$ and $j_2$. Each block of the tensor is, therefore, labeled by three spins,
\begin{equation*}
\diagram{main}{4},
\end{equation*}
and is only non-zero when $j_1$ and $s$ can fuse to $j_2$ according to the fusion rules for irreps of $\SU(2)$.
\par Let us first investigate what this implies for a spin-1/2 chain. Suppose we want to write down an $\SU(2)$ invariant MPS with a one-site unit cell. Since the MPS tensor $A$ has to transform as a singlet (i.e., according to the trivial representation), we have the following condition on the allowed irreps on the bonds
\begin{equation*}
\diagram{main}{5} \rightarrow \left| j_1 - j_2 \right| =\frac{1}{2},
\end{equation*}
which implies that a half-integer $j_1$ only couples to an integer $j_2$, and vice versa. If we build up an MPS using this tensor,
\begin{equation*}
\diagram{main}{6},
\end{equation*}
then this state falls apart into a sum of two states where the first state has $j_1$ half-integer, $j_2$ integer and so forth, and the second state contains the other representations. Therefore, for describing a singlet ground state for an $s=1/2$ spin chain, we need at least a two-site unit cell, where we alternate between half-integer and integer representations:
\begin{equation*}
\diagram{main}{7}
\end{equation*}
where $j_1$ ($j_2$) only has half-integer (integer) representations, or vice versa.
\par This result implies that an MPS cannot describe a unique ground state of a spin-1/2 chain, a result that is closely connected to the Lieb-Schultz-Mattis theorem \cite{Lieb1961, Oshikawa2000}, stating that a spin-1/2 chain cannot host a non-degenerate ground state with a gapped spectrum. Here we find that an MPS representation for a ground state necessarily breaks translation invariance, and, therefore that the translated state is an equally good ground-state approximation. The simplest example of an $\SU(2)$-invariant MPS on a spin-1/2 chain is the Majumdar-Ghosh state \cite{Majumdar1969a, Majumdar1969b}, which is obtained by interchanging $j=0$ and $j=\frac{1}{2}$ representations on the virtual bonds.
\par The situation for integer spin chains is very different. Haldane famously showed that spin-1 chains typically have a unique ground state with a finite excitation gap \cite{Haldane1983a, Haldane1983b}. Using the MPS framework, it was shown that spin-1 chains host symmetry-protected topological (SPT) phases \cite{Pollmann2010, Chen2011, Schuch2011} that are characterized by a string-order parameter, spin-1/2 edge states and a ground-state entanglement spectrum with even degeneracies. The transition to a trivial phase can only occur through a phase transition. Here, the simplest example is the Affleck-Kennedy-Lieb-Tasaki state \cite{Affleck1987, Affleck1988}, obtained by taking only $j=\frac{1}{2}$ representations on the bonds.
\par The characteristic difference between an SPT phase and a trivial phase is again clearly seen in the MPS description of the $\SU(2)$-invariant ground state. For a spin-1 chain, the MPS tensors necessarily have virtual representations $j_1$ and $j_2$ with the property
\begin{equation*}
\diagram{main}{8} \rightarrow \left| j_1 - j_2 \right| = 0,1.
\end{equation*}
This implies that $j_1$ and $j_2$ are either both half-integer or both integer. This implies that MPS representations of ground states are possible using a one-site unit cell
\begin{equation*}
\diagram{main}{9},
\end{equation*}
where all $j$'s are either integer or half-integer. These two cases differentiate a trivial from an SPT phase, respectively: The degeneracies in the entanglement spectrum are determined by the multiplicities ($m=2j+1$) of the $\SU(2)$ representations, such that integer ones correspond to odd degeneracies and the half-integer ones to even degeneracies.

\subsection{Elementary excitations}
\label{sec:spinon}

Besides ground states, the uniform MPS framework can be extended to the description of elementary excitations. Indeed, it was rigorously shown that an excitation that lives on an isolated branch in the spectrum can be described by acting with a momentum superposition of a local operator onto the ground state \cite{Haegeman2013}. In the MPS language, this translates to a quasiparticle ansatz for elementary excitations on top of an MPS ground state \cite{Haegeman2012}. When applied to an $\SU(2)$ invariant spin system with a unique translation-invariant ground state, we have the following form for an elementary excitation:
\begin{equation*}
\ket{\Phi_p^k(B)} = \sum_{n} \e^{ipn} \diagram{main}{10}.
\end{equation*}
Here we have introduced a new tensor $B$ that perturbs the ground state in a local region around site $n$, and performed a plane-wave superposition with momentum $p$. We have added an extra leg to this tensor that transforms according to a certain $\SU(2)$ irrep, labeled by $k$. Therefore, the irrep that lives on this non-contracted leg determines the global quantum number of the excited state.
\par The ansatz wavefunction is linear in the tensor $B$, and therefore the variational parameters in $B$ can be optimized by solving a generalized eigenvalue problem. Using a well-chosen parametrization for the tensor $B$, the norm of the wavefunction can be made trivial, which reduces the generalized eigenvalue problem to an ordinary one. When the quantum numbers of the excitation---the momentum $p$ and the $\SU(2)$ label $k$---are non-trivial, the excitation is orthogonal to the ground state by construction; for trivial quantum numbers the ansatz can be made orthogonal by the same well-chosen parametrization.
\par Evidently, when considering integer-spin chains, regardless of whether the $j$'s are integer or half-integer, the label $k$ has to be an integer. This corresponds to the well-known property that spin-1 chains generically have magnon excitations. 
\par In the half-integer spin case, where the MPS breaks translation invariance and has a two-site unit cell, we can make elementary excitations by considering defects in the ground-state pattern. Indeed, starting from an MPS ground state with tensors $A_1$ and $A_2$ 
\begin{equation*}
\ket{\Psi(A_1,A_2)} = \diagram{main}{11},
\end{equation*}
where $j_i$ ($j_h$) denote integer (half-integer) representations, an excitation would look like
\begin{equation*}
\ket{\Phi_p^k(B)} = \\ \sum_{\text{$n$ even}} \e^{ipn} \diagram{main}{12}.
\end{equation*}
One now observes that the label $k$ has to be half-integer, which indicates that elementary excitations in spin-1/2 chains generically have half-integer quantum numbers. These spinons cannot be created out of the ground state by a local operator, but are always created in pairs; this phenomenon is known as fractionalization.
\par From the MPS perspective, therefore, it is natural that half-integer spin chains host spinon excitations, whereas magnons appear in integer-spin chains. There is, however, a scenario conceivable where spinon excitations can emerge in a spin-1 chain. We imagine that we can find a hamiltonian for which we have two different MPS ground states $\ket{\Psi(A_1)}$ and $\ket{\Psi(A_2)}$ with exactly the same ground-state energy density, where one MPS carries only integer representations on the legs and the other only half-integer ones. Such a scenario typically occurs at a first-order transition between an SPT phase and a trivial phase. In that case, we can consider solitonic excitations between the two ground states, 
\begin{equation*}
\ket{\Phi_p^k(B)} = \\ \sum_{n} \e^{ipn} \diagram{main}{13}. 
\end{equation*}
Now it is obvious that the irrep label $k$ has to be half-integer, so it carries fractional quantum numbers.

\subsection{Spinon/anti-spinon bound states}
\label{sec:bound}

Spin chains that host spinon excitations can often be perturbed such that the spinons are confined. In the above case of spin-1/2 chains, the easiest option is to favour one of the two ground-state patterns through an explicit dimerization in the spin-chain hamiltonian. In that case, the spinons no longer exist as elementary excitations, but if the perturbation is weak, one can still understand the low-lying excitations as spinon/anti-spinon bound states. These can be pictured as consisting of two local defects in the ground state pattern, and can be described by a two-particle wavefunction of the form
\begin{multline*}
\ket{\Phi_p^k(B)} = \sum_{\text{$n$ even}} \e^{ipn} \sum_{n'>0, \mathrm{even}} c(n') \\ \diagram{main}{14},
\end{multline*}
where $c(n')$ is the part of the two-particle wavefunction for the relative position between the two spinons. 
\par In the case of spinons in a spin-1 chain at a first-order transition line, a spinon/anti-spinon wavefunction would look like
\begin{multline*}
\ket{\Phi_p^k(B)} = \sum_{n} \e^{ipn} \sum_{n'>0} c(n') \\ \diagram{main}{15}.
\end{multline*}
Here spinon confinement can be easily introduced by tuning slightly away from the first-order point such that one of the two ground states is favoured over the other energetically.
\par This type ansatz wavefunction was introduced for describing two-particle scattering states \cite{Vanderstraeten2014, Vanderstraeten2015}, for which the relative wavefunction $c(n')$ has an oscillating form. It was shown in Ref.~\onlinecite{Vanderstraeten2016} that the transition of a scattering state into a bound state corresponds to the relative wavefunction $c(n)$ changing from an oscillating function into an exponentially decaying one. This process of bound-state formation is signalled in the divergence of the scattering length, which can be read off from $c(n)$ \cite{Vanderstraeten2016}.
\par In principle, however, the description of stable bound states fall within the above one-particle framework: their wavefunctions are constructed as local deformations of the ground state in a momentum superposition. Indeed, for strongly-bound states, the one-particle ansatz has proven to be sufficient to capture the wavefunction accurately \cite{Vanderstraeten2016, Bera2017}. However, when very broad bound states are considered---when the two $B$ tensors are well separated---the above quasiparticle ansatz can be insufficient in the sense that a single local tensor cannot capture the full extension of the ground-state perturbation. In that case, an extended ansatz of the form \cite{Haegeman2013}
\begin{equation*}
\ket{\Phi_p^k(B)} = \\ \sum_{n} \e^{ipn}  \diagram{main}{16}
\end{equation*}
can be introduced. The number of parameters in the $B$ tensor scales exponentially with the number of sites $N$, such that a variational optimization becomes unfeasible rather quickly. For that reason, we can decompose the $B$ tensor in a string of $N$ one-site tensors giving rise to the ansatz
\begin{equation*}
\ket{\Phi_p^k(B)} = \\ \sum_{n} \e^{ipn} \diagram{main}{17}.
\end{equation*}
The variational optimization of the string of tensors can be performed using a sweeping algorithm, much in the spirit of standard DMRG \cite{Schollwoeck2011} algorithms---we refer to the appendix for more details on the implementation.

\section{Numerical study of spinons and their confinement in a spin-1 chain}
\label{sec:spin1}

In the first part of this paper, we have laid out the MPS formalism for capturing generic cases of spinons and their confinement in $\SU(2)$-invariant spin chains. In addition, we have proposed the scenario in which spin-1/2 spinons can emerge in a spin-1 chain on a first-order transition line. This phenomenon was observed to occur in a frustrated and dimerized spin-1 chain \cite{Chepiga2016}---the spin-1 chain with next-nearest neighbour and biquadratic interactions shows a similar phenomenology \cite{Pixley2014, Chepiga2016c}.
\par In the second part, we apply our formalism to study the spinon excitations in the former model numerically. In addition, we simulate the confinement of these spinons away from the transition line. Note that in Refs.~\onlinecite{Vanderstraeten2016, Bera2017} spinon confinement in spin-1/2 chains was already simulated using the framework of uniform MPS without symmetries. In the following we have performed the simulations using tangent-space methods for uniform MPS \cite{Vanderstraeten2019} using full $\SU(2)$-symmetric tensor-network operations \cite{McCulloch2002, Weichselbaum2012, symmTN}.

\begin{figure}
\includegraphics[width=0.8\columnwidth]{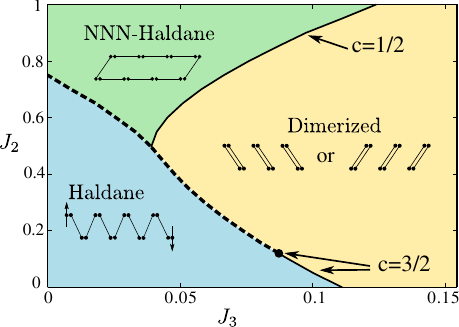}
\caption{The phase diagram of the frustrated and dimerized spin-1 chain, taken from Ref.~\onlinecite{Chepiga2016}. The three phases are pictorially represented by a picture of virtual spin-1/2 particles: the SPT phase is a Haldane phase, the trivial phase can be interpreted as a next-nearest-neighbour Haldane phase, and in the dimerized phase the pairing of the virtual particles leads to a spontaneous breaking of translation symmetry. The full (dashed) lines represent second-order (first-order) transitions.}
\label{fig:diagram}
\end{figure}

\par We investigate the frustrated and dimerized spin-1 chain, defined by the hamiltonian
\begin{multline*}
H = J_1 \sum_i  \vec{S}_i \cdot \vec{S}_{i+1} + J_2 \sum_i \vec{S}_{i-1} \cdot \vec{S}_{i+1}  \\ + J_3 \sum_i \left((\vec{S}_{i-1} \cdot \vec{S}_{i})(\vec{S}_{i} \cdot \vec{S}_{i+1} ) + h.c. \right).
\end{multline*}
For $J_2=J_3=0$ this model reduces to the spin-1 Heisenberg model, which is known to be in the Haldane phase \cite{Haldane1983a, Haldane1983b}. The next-nearest neighbour term ($J_2$) adds frustration to the system and drives it through a first-order phase transition into a trivial phase \cite{Kolezhuk1996, Kolezhuk1997}, whereas the three-site interaction ($J_3$) induces a spontaneous dimerization via a second-order transition \cite{Michaud2012}. The full phase diagram (see Fig.~\ref{fig:diagram}) shows that the first-order transition extends over a finite region, and only for small $J_2$ the transition into the dimerized phase becomes second order.

\subsection{Deconfined spinons between SPT and trivial phase}

\begin{figure}
\includegraphics[width=0.99\columnwidth]{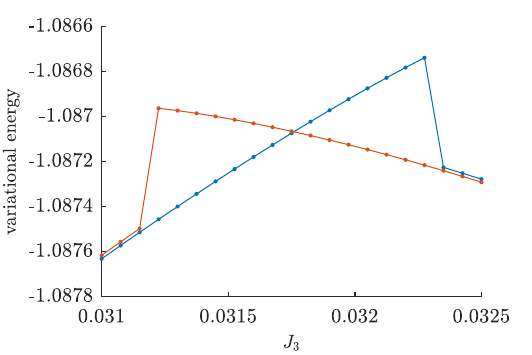}
\caption{The variational energy obtained by MPS with half-integer (blue) and integer (red) representations on the links for $J_1=1$, $J_2=0.56$, and as a function of $J_3$. We observe a crossing indicating a first-order transition between an SPT phase and a trivial phase. The blue (red) data points were obtained by taking the previous MPS as the starting point for the next point, such that the optimization algorithm stays in the local minimum corresponding to the higher-energy state. Ultimately, the variational optimization drops farther away from the phase transition, where the MPS with the wrong representations will develop a non-injective structure to approximate the true ground state.}
\label{fig:transition}
\end{figure}

\par Let us first investigate the first-order transition between the SPT phase and the trivial phase. On both sides of the transition, we can represent the ground state by an MPS with a one-site unit cell with an explicit encoding of the $\SU(2)$ symmetry: In the SPT phase, we impose half-integer representations on the virtual degrees of freedom, whereas in the trivial phase we impose only the integer ones. Because these two choices determine different classes of MPS, we can compare the variational energies within the two distinct classes and determine in which phase the ground state is for a given choice of parameters. In Fig.~\ref{fig:transition} we plot the variational energies on a line in the phase diagram that crosses the transition, showing nicely that this is, indeed, a first-order transition.

\begin{figure}
\subfigure[]{\includegraphics[width=0.99\columnwidth]{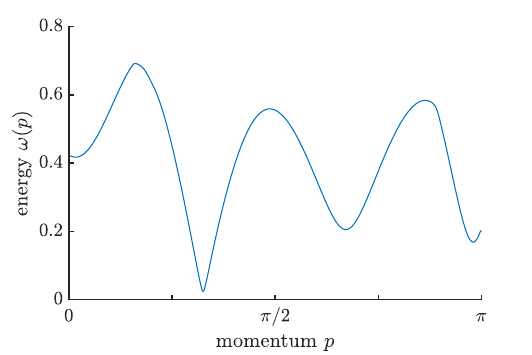}} \\
\subfigure[]{\includegraphics[width=0.49\columnwidth]{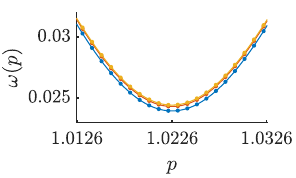}} 
\subfigure[]{\includegraphics[width=0.49\columnwidth]{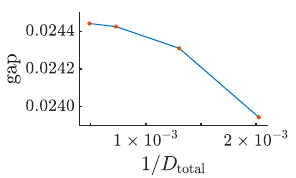}}
\caption{In (a) we plot the dispersion relation of the spinon excitation at the first-order phase transition between Haldane and trivial phase at $J_2=0.56$, $J_3\approx0.0318$. The full dispersion relation has been computed using the spinon quasiparticle ansatz with $\SU(2)$ symmetry. To get an idea of the convergence as a function of bond dimension, in (b) and (c) we plot the dispersion around the minimum and the convergence of the gap with higher bond dimensions. The three curves in (b) correspond to the three points with highest bond dimension in (c).  The highest bond dimension we used is $D=200$ for the largest subblock; for comparison, this corresponds to a non-symmetric MPS with total bond dimension $D_\mathrm{total}\approx2000$. Note that the excitation energy is not variational, because we subtract the MPS ground-state energy.}
\label{fig:spinon00318}
\end{figure}

\begin{figure}
\includegraphics[width=0.99\columnwidth]{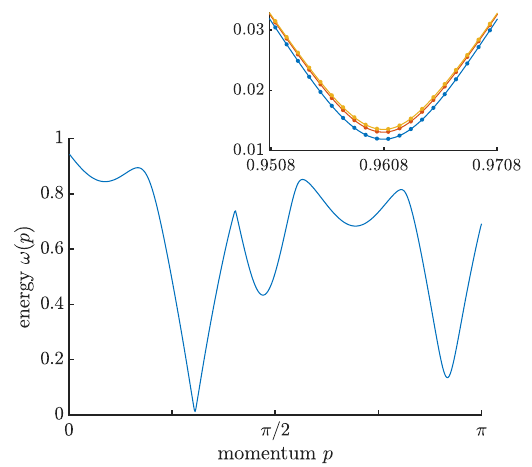}
\caption{The same plot as before, now for parameters $J_2\approx0.7606$ and $J_3=0$ (the bond dimensions for the different curves in the inset are the same as before). The inset shows that the spinon has decreased significantly with respect to the previous figure.}
\label{fig:spinon0}
\end{figure}

\par Exactly at the transition, the two ground states have the same energy density. Therefore, we can consider domain walls that interpolate between them, where, as we have shown in the previous section, the excitations necessarily carry a half-integer quantum number. Using the variational quasiparticle ansatz for spinons interpolating between two MPS ground states with half-integer and integer representations on the virtual bonds (see Sec.~\ref{sec:spinon}), we can compute the spinon excitation energy at all values of the momentum. In Fig.~\ref{fig:spinon00318} we plot our results for the dispersion relation of the spinon with fractional spin $s=\frac{1}{2}$ for $J_2=0.56$ and $J_3\approx0.0318$. The spinon's dispersion relation exhibits a very strong minimum at an incommensurate value of the momentum. In the inset, we provide a close-up around the minimum showing that the gap converges to a non-zero value. In addition, in Fig.~\ref{fig:spinon0} we plot the dispersion relation further along the transition line ($J_2\approx0.7606$, $J_3=0$) showing that the spinon gap decreases. It is expected that the gap ultimately closes when going further along this line towards negative $J_3$---this closing of the gap can be described by a marginal operator changing sign in the $\SU(2)_1$ Wess-Zumino-Witten field theory with central charge $c=1$ \cite{Chepiga2016b, Tsui2017}. A continuous transition with $c=1$ between an SPT chain and a trivial phase was recently demonstrated in Ref.~\onlinecite{Gozel2019}. Unfortunately, we have found no immediate evidence for a critical point further along the transition line, and we leave an elaborate study of this question for further work.

\begin{figure}
\includegraphics[width=0.99\columnwidth]{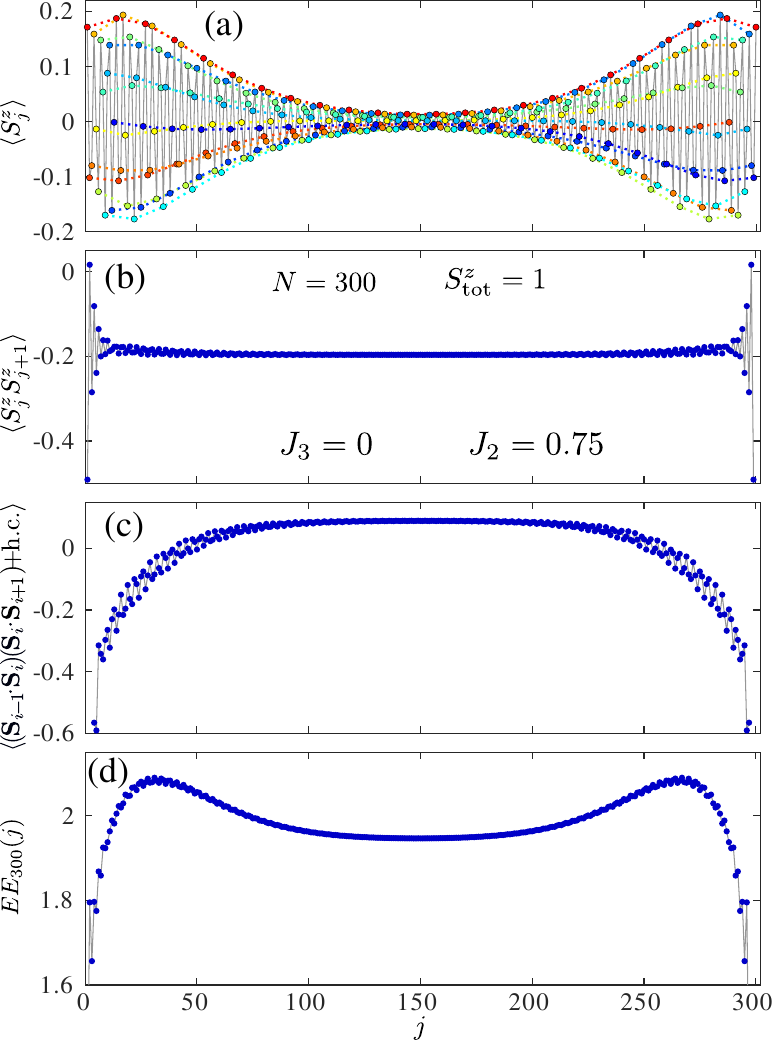}
\caption{(a) Local magnetization, (b) nearest-neighour correlations, (c) three-site correlations, and (d) bipartite entanglement profile for a chain with $N=300$ sites and $S^z_\mathrm{tot}=1$ at the first-order transition between the SPT and trivial phase at $J_2\approx0.75$ and $J_3=0$. The colors in (a) are a guide to the eye to show the periodicity in the correlations: we have used the same color for every 14th site.}
\label{fig:dmrg1}
\end{figure}

\par The existence of the spinons as low-energy excitations at the phase transitions can be further confirmed from simulations on a finite chain \cite{Chepiga2016}. Indeed, at the first-order transition the energy levels of the corresponding states cross and one can observe the coexistence of different domains on a finite chain. In Fig.~\ref{fig:dmrg1} we show the results at the first-order transition between the topologically trivial and SPT phases at $J_2=0.75$, $J_3\approx0$. Four quantities are most relevant: the local magnetization $\langle S_j^z\rangle$, the nearest-neighbor correlations, the three-site correlations, and the bipartite entanglement entropy $EE_N$. In the middle of the chain we can see the signature of the Haldane phase: negative nearest-neighbor correlations dominate, whereas the three-site correlations are positive (as the product of two nearest-neighbour terms). At the edges, the three-site correlations are strongly negative. From the entanglement entropy, which takes its maximal value where the spinon is situated, we can see that domain walls separating these different domains approximately 30 sites away from the edges. Although the local magnetization profile shown in Fig.~\ref{fig:dmrg1}(a) is significantly perturbed by incommensurate correlations, one can clearly see that the maximum of the amplitude is shifted away from the boundary.

\subsection{Confinement of spinons away from the transition line}

\begin{figure}[!htbp] \centering
\includegraphics[width=0.99\columnwidth]{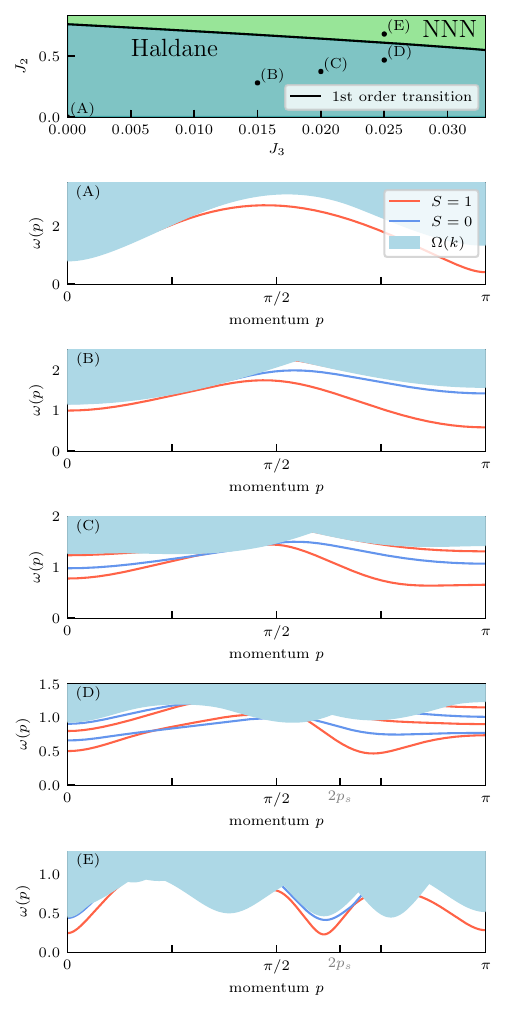}
\caption{Excitation spectra for different points in the phase diagram. The lines are the first low-lying spin-0 and spin-1 excitations, whereas the blue shaded areas are multi-particle continua computed from taking combinations of single-particle excitations and adding their momenta and energies. We observe an accumulation of bound-state modes in the spectrum as the first-order transition is approached. The third spectrum also shows the formation of an incommensurate minimum. The extra tick in grey shows twice the momentum for which the spinon dispersion relation is minimal, see Fig.~\ref{fig:spinon00318}. The simulations were performed at bond dimension $D_\mathrm{total}=120$.}
\label{fig:spectra}
\end{figure}

\begin{figure} \centering
\includegraphics[width=0.99\columnwidth]{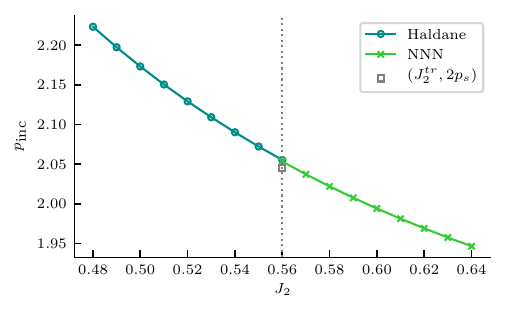}
\caption{The momentum for which the dispersion relation reaches its incommensurate minimum at $J_3 = 0.0318$ as a function of $J_2$. At $J_2 \approx 0.56$ the system undergoes a first-order phase transition. The value of $p_{\mathrm{inc}}$ is compatible with two times the momentum of the gap in the free spinon dispersion relation (see Fig.~\ref{fig:spinon00318}, $p_s \approx 1.0226$) as indicated by the grey square. The simulations were performed at bond dimension $D_\mathrm{total}=120$.}
\label{fig:2ps_minima}
\end{figure}

The spinons that we have identified in the previous section exist as freely propagating particles only at the first-order phase transition exactly, but their existence has noticeable effect away from the transition as well. Indeed, we have observed in Fig.~\ref{fig:transition} that both ground states still exist independently away from the transition point, where one of the two will have slightly lower energy density. As we have explained in Sec.~\ref{sec:bound}, we can still consider spinon/anti-spinon pairs against the background of the ground state that is favoured energetically. The excess of energy between the spinon and anti-spinon due to the higher-energy background state between them causes the spinon/anti-spinon pair to experience a linear potential and form bound states. As discussed in the introduction, this phenomenon has been studied extensively in spin-1/2 chains \cite{Sorensen1998, Affleck1998, Augier1999, Shiba1980, Vanderstraeten2016, Bera2017}.
\par The formation of spinon/anti-spinon bound states away from the first-order phase transition is observed when plotting the excitation spectrum for a few values of the coupling inside the Haldane phase, see Fig.~\ref{fig:spectra}. Indeed, for $J_2=J_3=0$ we find the usual spectrum of the Heisenberg chain with a minimum in the dispersion at momentum $\pi$. When we come closer to the phase transition, we find that the minimum starts shifting to an incommensurate value, an observation that was also made from the real-space correlation functions \cite{Chepiga2016}. More interestingly, we find different isolated lines below the continuum emerging when approaching the phase transition. The minima of these isolated lines are situated at momentum $p=0$ and at an incommensurate value $p=p_{\mathrm{inc}}$. Above we have seen that the spinon dispersion relation has a strong minimum at momenta $\pm p_s$, and we expect to see spinon/anti-spinon bound states around the momenta that can be obtained by adding the individual momenta of the (anti-) spinons, i.e. around $p=p_s\pm p_s$.
\par In Fig.~\ref{fig:2ps_minima} we have explicitly tracked the location of these incommensurate minima. We varied $J_2$ at constant $J_3$ towards the first order transition point. From both the Haldane phase and the NNN-trivial phase, we indeed observe convergence towards $p=2p_s$ which confirms the confinement of the spinons away from the transition line.  

\begin{figure} \centering
\includegraphics[width=0.99\columnwidth]{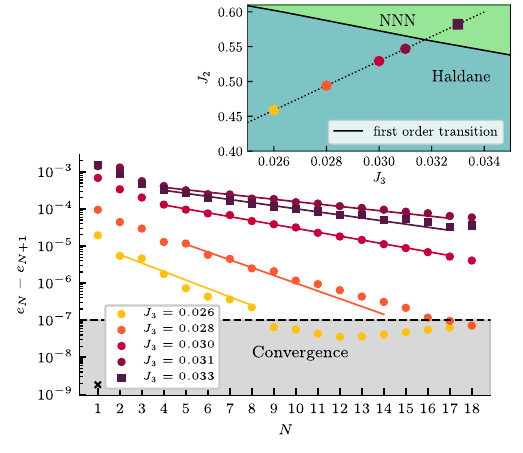}
\caption{The convergence of the variational excitation energies at momentum $k=0$ as a function of the spatial support $N$ of the string ansatz at various points near the phase transition (bottom). These points are located on the line through the origin and the transition point $J_2 = 0.56, J_3\approx  0.0318$ (top). The convergence becomes better further away from the phase transition, which is consistent with the bound state picture. For comparison we show the convergence of the magnon (with $k=\pi$) at the Heisenberg point $J_2=0,J_3 = 0$ (black cross).} \label{fig:conv}
\end{figure}

\par In addition we have applied the extended quasiparticle ansatz for broad bound states, containing a string of tensors [Sec.~\ref{sec:bound}]. In Fig.~\ref{fig:conv} we show the performance of this extended ansatz for the lowest-lying excitation in the system upon approaching the phase transition. When far away from the transition, the variational energy converges very quickly, which shows that the bound state has a limited spatial extent. If the transition is approached, the convergence becomes slower, which points to a widening of the spinon/anti-spinon bound state. The fact that the excitations become broad, extended perturbations of the ground state as the first-order transition is approached, confirms our underlying spinon picture for the low-lying excitations in the vicinity of the first-order phase transition.

\subsection{Spinons between SPT and dimerized phase}

\par We have also studied the first-order transition between the SPT phase and the dimerized phase for smaller $J_2$. The situation is slightly more complicated, because the dimerized phase itself hosts spinon excitations as well. Indeed, the dimerized phase has an MPS ground-state description with a two-tensor unit cell, and the low-energy particles are spin-1 defects in the ground-state pattern. In order to focus on the spin-1/2 spinons around the phase transition, we perform a blocking transformation such that a dimerized ground state maps to a translation-invariant MPS described by a single tensor with integer $\SU(2)$ representations on the virtual bonds. In this setting, the description of the spinons on the first-order transition line is similar as before.
\par For this case, it is known \cite{Chepiga2016b} that the first-order line ends and becomes second order, where the transition is described by a $\SU(2)_2$ Wess-Zumino Witten field theory with central charge c=3/2. The transition between first and second order---i.e., the closing of the spinon gap as one travels on the phase-transition line---is described by a marginal operator in the field theory changing sign.
\par
\begin{figure}
\includegraphics[width=0.99\columnwidth]{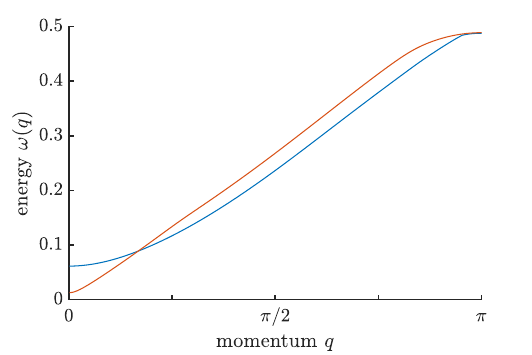}
\caption{The spinon dispersion relation on the first-order transition line between the SPT phase and the dimerized phase, for parameters $(J_2,J_3)$ given by $(0.3265,0.0558)$ (blue) and $(0.2915,0.0603)$ (red). The dimerized ground state breaks one-site translation invariance spontaneously, so the momentum $q$ is defined with respect to translations over two sites.}
\label{fig:disp3}
\end{figure}

 In Fig.~\ref{fig:disp3} we plot the spinon dispersion relation for two points on the first-order transition line between SPT and dimerized phase. First we observe that the minimum of the dispersion relation is at momentum $q=0$, so we do not have any commensurate correlations in the system. Moreover, we observe that the spinon gap becomes smaller quickly as we travel on the transition line towards the critical point. This rapid decrease of the gap is expected from the field-theory description \cite{Chepiga2016b}, which predicts an exponential suppression as the critical point is approached.

\begin{figure}
\includegraphics[width=0.99\columnwidth]{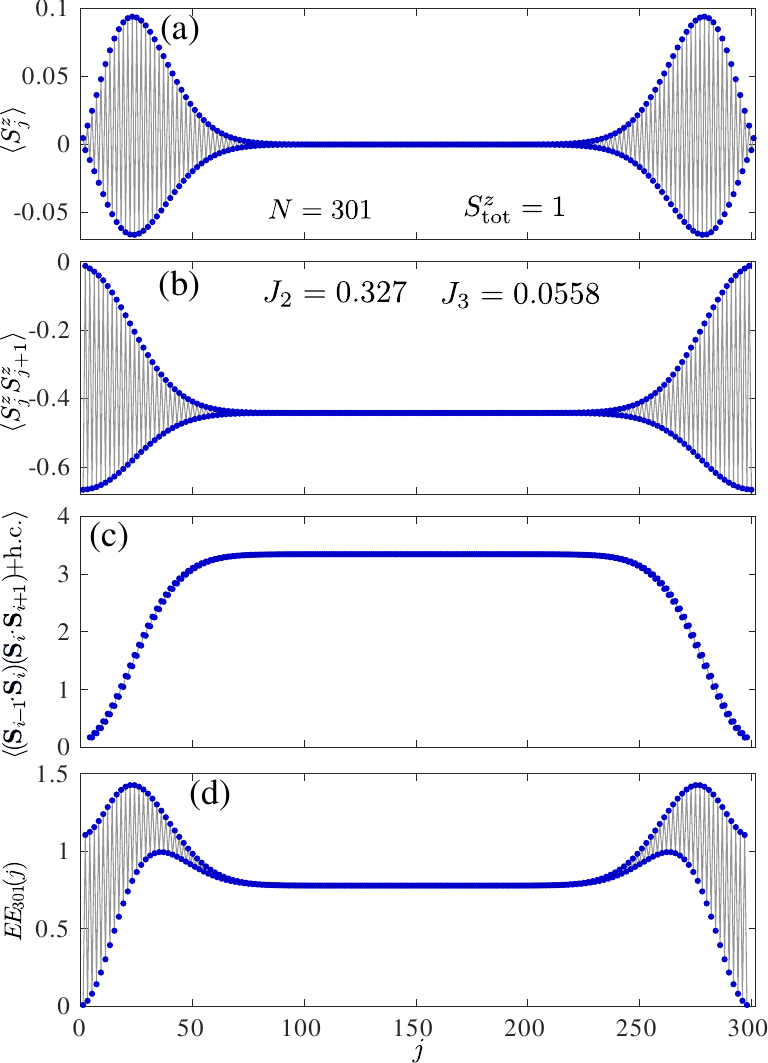}
\caption{(a) Local magnetization, (b) nearest-neighour correlations, (c) three-site correlations, and (d) bipartite entanglement profile for a chain of $N=301$ sites and $S^z_\mathrm{tot}=1$ at the first-order transition between the SPT and dimerized phase at $J_2=0.327$ and $J_3=0.0558$.}
\label{fig:dmrg2}
\end{figure}

\par In the present case we also confirm the existence of the spinons by looking at the finite-size profiles listed above. Based on Figs.~\ref{fig:dmrg2}(a), (b) and (c) one can deduce that open edges favor dimerized domains, while the central region remains in the SPT phase. Indeed, here the Haldane phase is signaled by the negative value of the nearest-neighbour correlations and the large positive value for the three-site correlations, whereas the nearest-neighbour correlations strongly oscillate for the dimerized phase. Note, however, that for a selected coupling constant the SPT domain is still commensurate, which implies that the ground-state is a singlet if the total number of sites is even, and the ground-states is a (Kennedy) triplet \cite{Kennedy1990}, if the total size of the domain is odd. Moreover, the dimerized domains necessary contain an even number of sites. So, keeping the total number of sites odd, we ensure a Kennedy triplet on the central SPT domain. According to Fig.~\ref{fig:dmrg2}(a) the domain walls are located at a distance about 25 spins from each edge, and the entanglement entropy also takes its maximal values at these locations.

\section{Conclusions}

To summarize, we have shown that $\SU(2)$ invariant spin-1 chains can support deconfined spinons at the first-order transition between the SPT Haldane phase and a topologically trivial phase without fractional edge excitations by actually calculating the spinon dispersion with an MPS ansatz along two transition lines with the NNN Haldane phase and the spontaneously dimerized phase respectively. The characterization of these excitations as deconfined spinons is further supported by the demonstration that they actually confine away from the transition line, as explicitly demonstrated in the case of the transition to the NNN Haldane phases for which the presence of dispersive bound states has been demonstrated on both sides of the transition.

Let us emphasize that the description of the bulk excitations of a spin-1 chain in terms of fractional spin-1/2 excitations is by no means straightforward since the basic local operators are spin-1 operators. This requires a description that simultaneously allows fractional excitations to delocalize and enforces the local projection on spin-1 states. Building on previous results on the description of elementary excitations with MPS, we have shown how spinons emerge naturally from the symmetry pattern of the ground-state MPS tensors for $\SU(2)$ systems, including integer spins. 

These results could be extented in several ways. For instance, the same analysis can be extended to chains with other global symmetries, where the case of $\SU(N)$ is arguably the most interesting. Indeed, a three-tensor ground state naturally appears for a $\SU(3)$ chain in the fundamental representation, and one can consider two types of defects or spinons against this background.

\par In two dimensions, we already know that the framework of projected entangled-pair states (PEPS) allows us to simulate even more exotic quasiparticles \cite{Vanderstraeten2015b, Vanderstraeten2019b}. Indeed, whereas the one-dimensional case only allows for defects in the ground-state pattern, in two dimensions we can consider quasiparticles with non-trivial strings of symmetry operations as well \cite{Schuch2011, Haegeman2015}. An $\SU(2)$-symmetric PEPS both hosts spinons and visons as elementary excitations \cite{Poilblanc2012}, and it would be interesting to study these quasiparticles and their confinement for spin-liquid hamiltonians.

\par As far as experiments are concerned, spinon excitations have been observed in neutron-scattering experiments on quasi-one-dimensional spin-1/2 compounds \cite{Nagler1991, Tennant1993, Arai1996, Mourigal2013, Lake2013}. The fact that the spinons necessarily come in pairs leads to a broad continuum in the dynamical structure factor, in contrast to the more conventional magnon mode. In more recent neutron-scattering experiments, the confinement of these particles has been observed by the splitting of the multi-spinon continuum into a stack of bound states \cite{Grenier2015, Bera2017}. From that point of view, the main message of the present paper is to argue that a similar transition between confined and deconfined spinons is present in frustrated spin-1 chains. All it takes is a significant next-nearest coupling typical of zig-zag chains to induce a first-order transition between the Haldane and the next-nearest neighbor Haldane phases. We hope that the present paper will stimulate the search for such spin-1 chains and their experimental investigation with inelastic neutron scattering.

\par We acknowledge insightful discussions with Ian Affleck about solitons in spin chains. This research is supported by the Research Foundation Flanders (LV, FV), the Swiss National Science Foundation (NC, FM), the German Research Foundation (Deutsche Forschungsgemeinschaft, DFG) through TRR80 Project F8 (EW), and ERC grant QUTE (FV). 

\bibliography{./bibliography}

\appendix

\input{appendix_string}

\end{document}

%% file: preamble.tex
\usepackage{graphicx}
\usepackage{graphics}
\usepackage{amsmath}
\usepackage{amssymb}
\usepackage{amsfonts}
\usepackage{dsfont}
\usepackage{braket}
\usepackage{color}
\usepackage{braket,slashed}
\usepackage[mathscr]{euscript}
\definecolor{darkblue}{rgb}{0, 0, 0.8}
\usepackage[colorlinks=true, breaklinks=true, linkcolor=red, citecolor=blue, urlcolor=blue]{hyperref} 
\usepackage{hyperref}
\usepackage{subfigure}
\usepackage{xfrac}
\usepackage{bm}
\usepackage{kantlipsum}
\usepackage{enumitem}
\usepackage{tikz}
\usepackage{framed}
\usepackage{graphicx}
\usepackage{subfigure}
\usepackage{booktabs}

\newcommand{\diagram}[2]{\;\vcenter{\hbox{\includegraphics[scale=0.35,page=#2]{./diagram_#1.pdf}}}\;}

\allowdisplaybreaks[1]

\newcommand{\code}[1]{\texttt{#1}}

\newcommand{\SU}{\ensuremath{\mathrm{SU}}}

\newcommand{\one}{\mathds{1}}

\renewcommand{\dag}{^\dagger}

\newcommand{\e}{\ensuremath{\mathrm{e}}}

%% file: appendix_string.tex
\section*{Appendix: String ansatz for broad excitations}

\newcommand{\rb}[1]{|#1)}
\newcommand{\lb}[1]{(#1|}
\newcommand{\ew}[1]{[\textcolor{red}{\textbf{EW:}} #1]}

In this appendix we provide the details of the extended quasiparticle ansatz for describing broad low-energy excitations (see Sec.~\ref{sec:bound} in main text). 

\subsection{Implementation}

We first recall that the quasiparticle ansatz is a momentum superposition of the uniform MPS ground state in which one tensor is distorted  
\begin{equation} \label{eq:qp}
\ket{\Phi_p(B)}= \sum_n e^{ipn} \diagram{appendix}{1} .
\end{equation}
For further details on uniform MPS, and in particular of the gauge fixing, we refer to the tangent-space review in Ref.~\onlinecite{Vanderstraeten2019}. Here we just mention that we always work in the mixed gauge and that the distortion tensor $B$ obeys the left gauge-fixing condition $\sum_{s=1}^d A_L^{s \dag} B^s = \sum_{s=1}^d B^{s \dag}  A_L^{s} = 0$
in which we assumed the ground-state tensor $A$ to be in the left-canonical form $\sum_s A_L^{s\dag} A_L^{s} = \one$. 
\par The quasiparticle ansatz~\eqref{eq:qp} describes local and low-energy excitations to extreme precision \cite{Haegeman2011}. However, as the variational subspace on top of the MPS ground state is very localized, it is not expected to accurately capture the effect of large physical operators acting on the ground state~\cite{Haegeman2013}. Therefore, broad excitations such as the spinon/anti-spinon bound states in the spin-1 chain discussed in the main text are not well described by this ansatz. 
\par In Ref.~\onlinecite{Haegeman2013} it is suggested to increase the variational support by spreading the distortion over several sites 
\begin{multline}\label{eq:block_qp}
\ket{\Phi_p(B)}=\\ \sum_n e^{ipn} \diagram{appendix}{2}.
\end{multline}
The $B$-block in this ansatz contains $D^2 d^N$ elements, where $D$ is the bond dimension and $d$ the physical dimension. By taking into account the gauge fixing $D^2 (d-1)d^{N-1}$ elements are truly variational. The exponenial scaling in the number of distorted sites makes this ansatz hard to use, except for the paradigmatic AKLT-model~\cite{Haegeman2013, Haegeman2013b}, for which the ground state can be exactly represented by an MPS with bond dimension 2. 
\par A more efficient representation is given by a tensor decomposition of the $B$-block
\begin{multline} \label{eq:string_qp}
\ket{\Phi_p(B_1 \cdots B_N )} =  \\ \sum_n e^{ipn} \diagram{appendix}{3}. 
\end{multline}
Here the gauge fixing condition only applies on first tensor ($B_1$), such that the other tensors in the decomposition ($B_2,\dots,B_N$) are purely variational.
\par By this decomposition the number of variational parameters can be chosen to scale linear with $N$. This clearly depends on the choice of limiting bond dimension $D_{\max}$ inside the excitation string. Now the excitation string can be seen as a finite-size subsystem on top of the ground state, and can be optimized by standard finite-size DMRG-methods \cite{Schollwoeck2011} that are computationally efficient. 
\par In order to construct such an efficient DMRG scheme to optimize over the excitation tensors $B^{s_{n}}_1 \cdots B^{s_{n+N-1}}_N$ in~\eqref{eq:string_qp}, we need to construct the effective Hamiltonian for a one-site update  
\begin{multline} \label{eq:Heff_mat}
2 \pi \delta(0) \bm{B}_i^{\dag} H_\mathrm{eff}^i(p)\bm{B}_i = \\ \braket{\Phi_p(B_1 \dots B_N ) | \hat{H} | \Phi_p(B_1 \dots B_N )},
\end{multline}
where the bold symbols denote the vectorized version of the corresponding tensor, and $\hat{H} = \sum_n \hat{h}_{n,\dots,n+M-1}$ the many-body Hamiltonian that only consists of local $M$-body interactions. Two-site update schemes may be equally well considered, but this will make the construction of the effective Hamiltonians more cumbersome. The construction of $H_\mathrm{eff}^i(p)$ boils down to the knowledge of the matrix element at the right hand side of Eq.~\eqref{eq:Heff_mat}. Translation invariance implies that the terms in the matrix element contain maximally a double infinite sum over transfer matrices $ \sum_{s=1}^d A^s \otimes \bar{A}^{s}$. But still we need to take into account all different relative positions of the local Hamiltonian $\hat{h}$ with respect to all excitations tensors that appear in the bra and in the ket-layer. Here the left gauge fixing of the first excitation tensor significantly reduces the number of terms.
\par Once we have constructed $H_\mathrm{eff}^i(p)$, we can update the $i$-th site by solving the generalized eigenvalue problem 
\begin{equation} \label{eq:string_effeig}
H_\mathrm{eff}^i(p)\bm{X}_i = \omega(p) N_\mathrm{eff}^i(p) \bm{X}_i .
\end{equation}
If the MPS ground state and the $B$-tensors are in the mixed canonical form at each update, the eigenvalue problem reduces to a standard problem, i.e. $N_\mathrm{eff}^i(p)$ is the unit matrix. By sweeping through the excitation string, the excitation energy is gradually lowered up to convergence. With the obvious initialization of $B_2,\dots,B_N= A$, the starting energy will be equal to the lowest energy obtained by ansatz~\eqref{eq:qp}. Higher energy excitations may be found by projecting away the lower-lying excitations.

\subsection{Benchmarks}

We demonstrate the accuracy of the string-ansatz by comparing its energy solution with the solution of the full problem given by Eq.~\eqref{eq:block_qp} for the fundamental magnon at momentum $p=\pi$ in the AKLT model. This comparison is shown in Tab.~\ref{tab:aklt}. We raise the bond dimension inside the excitation string up to the limiting values $D_\mathrm{lim}=54$ and $D_\mathrm{lim}=108$, this corresponds to an exact decomposition of the full $B$-tensor in terms of separate blocks up to respectively $N=7$ and $N=8$ sites. For $N>\left\lbrace 7,8\right\rbrace$ the number of variational parameters scales linearly in the number of added sites, instead of exponentially in ansatz~\eqref{eq:block_qp}. For $N>\left\lbrace 7,8\right\rbrace$ we can never recover the same precision as the original results in the first column, though the difference seems to be negligible in practice. However, because of the computational efficiency, we can go to a higher number of sites. Consequently, we can recover and even slightly improve the precision obtained by optimizing Eq.~\eqref{eq:block_qp}. 

\begin{table}
\begin{center}
\begin{tabular}{cccc}
\toprule
$N$ & tensor & string $D_{lim}=54$ &  string $D_{lim}=108$\\
 \midrule
1 &  .3703703703703 & .370370370370370 & .370370370370370\\
2 &  .3506345810861 & .350634581086136 & .350634581086135\\
3 &  .3501652022172 & .350165202217295 & .350165202217298\\
4 &  .3501291730768 & .350129173076821 & .350129173076823\\
5 &  .3501247689418 & .350124768941853 & .350124768941852\\
6 &  .3501242254394 & .350124225439428 & .350124225439427\\
7 &  .3501241645674 & .350124164567495 & .350124164567493\\
8 &  .3501241580969 & .350124158096968 & .350124158096949\\
9 &  .3501241574175 & .350124157417571 & .350124157417519\\
10 & .3501241573460 & .350124157346082 & .350124157346044\\
11 & .3501241573384 & .350124157338518 & .350124157338485\\
12 & .3501241573376 & .350124157337713 & .350124157337683\\
13 & - & .350124157337627 & .350124157337597\\
14 & - & .350124157337619 & .350124157337586\\
 \bottomrule
\end{tabular}
\caption{Excitation energies of the magnon branch at momentum $\pi$ of the AKLT model. The ansatz substitutes $N$ sites in the MPS. The first column is copied from~\onlinecite{Haegeman2013} (see table on pag.~34) and is obained by the ansatz~\eqref{eq:block_qp}. The second and third column are obtained by the ansatz~\eqref{eq:string_qp} in which the internal bond dimension of the string is limited to $D_{lim}=54$ and $D_{lim}=108$ respectively. The eigensolvers used to obtain these energies are reliable up to 14 digits (15 digits are shown in the second and third column, and 13 in the first).} \label{tab:aklt}
\end{center}
\end{table}

We now consider the Ising model $S=1/2$ in a tilted magnetic field, the Hamiltonian is given by 
\begin{equation} \label{eq:IsSkew}
\hat{H} = - \sum_{i} \left(  \hat{\sigma}^x_{i}\hat{\sigma}^x_{i+1} + h_{\perp}  \hat{\sigma}^z_i + h_{\parallel}  \hat{\sigma}^x_i \right) , 
\end{equation}
the parameter $h_{\perp}$ describes a transverse field and the parameter $h_{\parallel}$ an additional longitudinal field. 

\begin{figure}
\centering
\includegraphics[width=0.99\columnwidth]{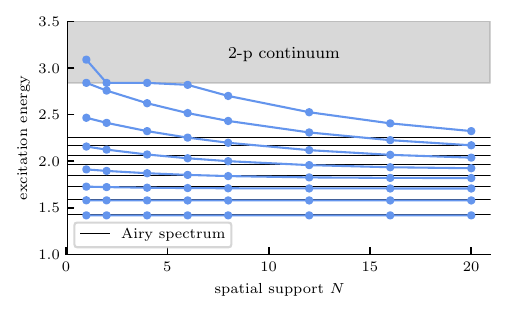}
\caption{Excitation energies at $p=0$ in the Ising model~\eqref{eq:IsSkew} for $h_{\perp}=0.7$ and $h_{\parallel}=0.0075$ as a function of the spatial support of the ansatz with $D_{lim}=40$. By increasing $N$ it becomes clear that the energies follow an Airy-like spectrum.   }\label{fig:airyspec}
\end{figure}

For $h_{\parallel} =0$ in the ordered regime far enough from the critical point, topological excitations as domain walls may occur. By applying the Jordan-Wigner transformation on the Hamiltonian, these domain walls can be represented as free fermions~\cite{Rutkevich2008}. 
When a longitudinal field $h_{\parallel}>0$ is applied, the $\mathbb{Z}_2$ symmetry is broken. This energetically favors one of the two previously degenerate ground states, and induces an attractive force between pairs domain walls -- they form a state of bound spinons. When the applied field is not too large, the force can be modeled by the cost of adding one site that is in the `wrong' ground state~\cite{Rutkevich2008}: $\mu = 2 h_{\parallel} \bar{m}$ where $\bar{m} = (1-h_{\perp}^2)^{1/8}$. Hence, the semi-classical Hamiltonian of the relative variables that describe the weakly confined spinons (or the slightly interacting fermions) just describes a particle that is moving in a linear potential. The time-independent Schr\"{odinger} equation for this Hamiltonian is the Airy equation. The low-lying energy spectrum is then approximated by the negative zero's of the anti-symmetric Airy function~\cite{Rutkevich2008}. At momentum $p=0$ the energy can be approximated as
\begin{equation} \label{eq:airyspec0}
E_n(0) \approx 4 (1-h_{\perp}) + \mu^{2/3} \left[ \frac{2 h_{\perp}}{1-h_{\perp}} \right]^{1/3}\xi_n.
\end{equation}
with $\xi_n$ determined by $\mathrm{-Ai}(-\xi_n)=0$ . 
We applied the ansatz~\eqref{eq:qp} and~\eqref{eq:string_qp} at $p=0$ to calculate the excitations in this model for $h_{\perp}=0.7$ in the weak confining regime with $h_{\parallel}=0.0075$. The results are shown in Fig.~\ref{fig:airyspec} together with the energies predicted by Eq.~\eqref{eq:airyspec0}. The quasiparticle ansatz ($N=1$) does not yet reveal the full Airy behavior of the spectrum. By increasing the spatial support of the excitation ansatz, we however observe a fast decrease of the excitation energies. The higher the excitation energy, the more significant the decrease of the energy. The highest excitation under study remains stuck in the continuum for the smallest $N$. The observation that the energies are always lower than the ones predicted from Eq.~\eqref{eq:airyspec0}, has probably to do with the effect of higher order terms in the expansion of the kinetic energy of the spinons. By going to a strong longitudinal field, we expect faster convergence as a function of $N$ but however stronger deviations from the Airy spectrum.

